\documentstyle[aaspp4,12pt]{article} 
\received{} 
\accepted{} 

\def\ltsima{$\; \buildrel < \over \sim \;$} 
\def\simlt{\lower.5ex\hbox{\ltsima}}            
\def\gtsima{$\; \buildrel > \over \sim \;$} 
 
\def\simgt{\lower.5ex\hbox{\gtsima}}            

\def\ha{{\sc H}$\alpha$} 
\def\nii{{\sc [Nii]}$\lambda\lambda$6548,6583\/} 
\def\niis{{\sc [Nii]}$\lambda$6583\/} 
\def\sii{{\sc [Sii]}$\lambda\lambda$6716,6730\/}

\def\cm3{cm$^{-3}$\/} 
\def\hb{{\sc{H}}$\beta$\/}

\def\oiii{{\sc{[Oiii]}}$\lambda$4959,5007\/} 
\def\o4363{{\sc{[Oiii]}}$\lambda$4363\/} 
\def\o4959{{\sc{Oiii]}}$\lambda$4959\/} 
\def\hii{{\sc{Hii}}\/} 
\def\hi{{\sc{Hi}}\/}

\def\heii{{\sc{Heii}}$\lambda$4686\/} 
\def\oiiis{{\sc [Oiii]}$\lambda$5007} 
 
\def\kms{km~s$^{-1}$} 
\begin{document} 
\title{\sc UGC~3995: A Close  Pair of Spiral Galaxies\altaffilmark{1}} 
\author{P. Marziani} 
\affil{Osservatorio Astronomico di Padova, Padova, Italia\\ e-mail: marziani@pd.astro.it} 
\author{M. D' Onofrio} 
\affil{Dipartimento di Astronomia, Universit\`a di Padova, Padova, Italia\\ 
e-mail: donofrio@pd.astro.it} 
\author{D. Dultzin-Hacyan} 
\affil{Instituto de Astronom\'{\i}a, UNAM, Aptdo. Postal 70-264,  M\'exico,   
D. F. 04510, M\'exico\\e-mail:deborah@astroscu.unam.mx} 
\author{J. W. Sulentic} 
\affil{Department of Physics \&\ Astronomy, University of Alabama, Tuscaloosa, 
USA\\e-mail:giacomo@merlot.astr.ua.edu} 
\altaffiltext{1}{Based on Observations collected at San Pedro Martir, Baja 
California, M\'exico.} 
\begin{abstract} 
 
UGC~3995 is a close pair of spiral galaxies whose eastern component hosts a 
Seyfert  2 nucleus. The object was selected because  a bright filamentary 
structure that apparently connects the nuclei of the two  galaxies made it a 
good candidate to investigate a possible interaction--AGN connection. We 
present a detailed  analysis of this system using long slit spectroscopy  and 
narrow (\ha\ + \nii) as well as broad band (B, R)  imaging and an archive  
WFPC2 image.  The component galaxies reveal surprisingly small signs of 
interaction considering their spatial proximity and almost identical recession  
velocities, as the bright filament  is probably an optical illusion due to the 
superposition of the bar of the Seyfert galaxy and of the spiral arms of the 
companion. 
 
The broad band morphology, a B--R  color map, and a continuum-subtracted \ha\ + 
\nii\ image demonstrate  that the western component UGC 3995B is in front of 
the Seyfert-hosting component UGC3995A, partly obscuring its western side. The 
small radial velocity difference leaves the relative motion of the two galaxies 
largely unconstrained. The observed lack of major tidal deformations, along
with some morphological peculiarities, suggests 
that the galaxies are proximate in space but may have 
recently approached each other on the plane of the sky. The geometry of the 
system and the radial velocity curve at P. A. $\approx$ 106$^\circ$\ suggest 
that the encounter may be retrograde or, alternatively, prograde before 
perigalacticon. 

The partial overlap of the two galaxies allows us to estimate the optical 
thickness of the disk of component B. We derive an extinction  $\approx$ 0.18 
visual magnitudes in the intra-arms parts of the foreground galaxy disk, and
$\simgt 1-1.5$  visual magnitudes in correspondence of the spiral arms. 
 
\end{abstract} 
 
 
\section{Introduction} 
 
Gravitational interactions have been proposed as a triggering mechanism
for Seyfert-type  activity since the early 70's. The interaction-AGN
connection is now part of the standard  paradigm of active galactic
nuclei (AGN; see
 \cite{iau186int}, \cite{stock99}, \cite{ost93}, \cite{der98}, \cite{bland90}, 
 for recent discussions emerging from different perspectives).
Gravitational disturbances are expected to be effective in transferring
angular momentum on scales of hundreds of parsecs (e. g. \cite{beg94},
and references therein), even though observational tests have produced
somewhat sparse or mixed results.   The AGN-interaction connection has
been investigated following two main observational approaches:  {\em
statistical}: counts of  the number of galaxies near to galaxies
hosting Seyfert nuclei and  {\em specific}: studies of the kinematics
and dynamics of ionized gas in spiral disks and circumnuclear regions
of individual Seyferts. Both methods have drawbacks:  present-day
observational techniques are not ideal for large scale studies that are
needed for a proper survey of companions within $\sim $ 1 Mpc, and  we
are unable to resolve structure within which nuclear activity seems to
be confined ($\simlt$ 1 pc). Several statistical papers based on sky
surveys have added important insight, even if some of them obtained
contradictory results (\cite{fins95};  \cite{iau186int} analyze the
origins of such disagreement). The most firm result is probably a
significant excess of bright companions in the vicinity of Seyfert 2
galaxies (\cite{deb98a}, \cite{iau186int}, \cite{fins95}). Restricting
the attention to dynamical studies of disk gas, direct observational
evidence in favor of interaction-driven neutral, molecular or ionized
gas infall toward the nucleus on scale of several hundreds of parsecs
remains scant.  Successful observations require  a special time in the
evolution of an interacting system, and detailed observations of  gas
morphology and kinematics. Perhaps only in the triple system NGC 7592
spectroscopic evidence of ionized gas infalling toward the Seyfert
nucleus has been convincingly found (\cite{pirlaf1}).

\begin{figure} 
\plotone{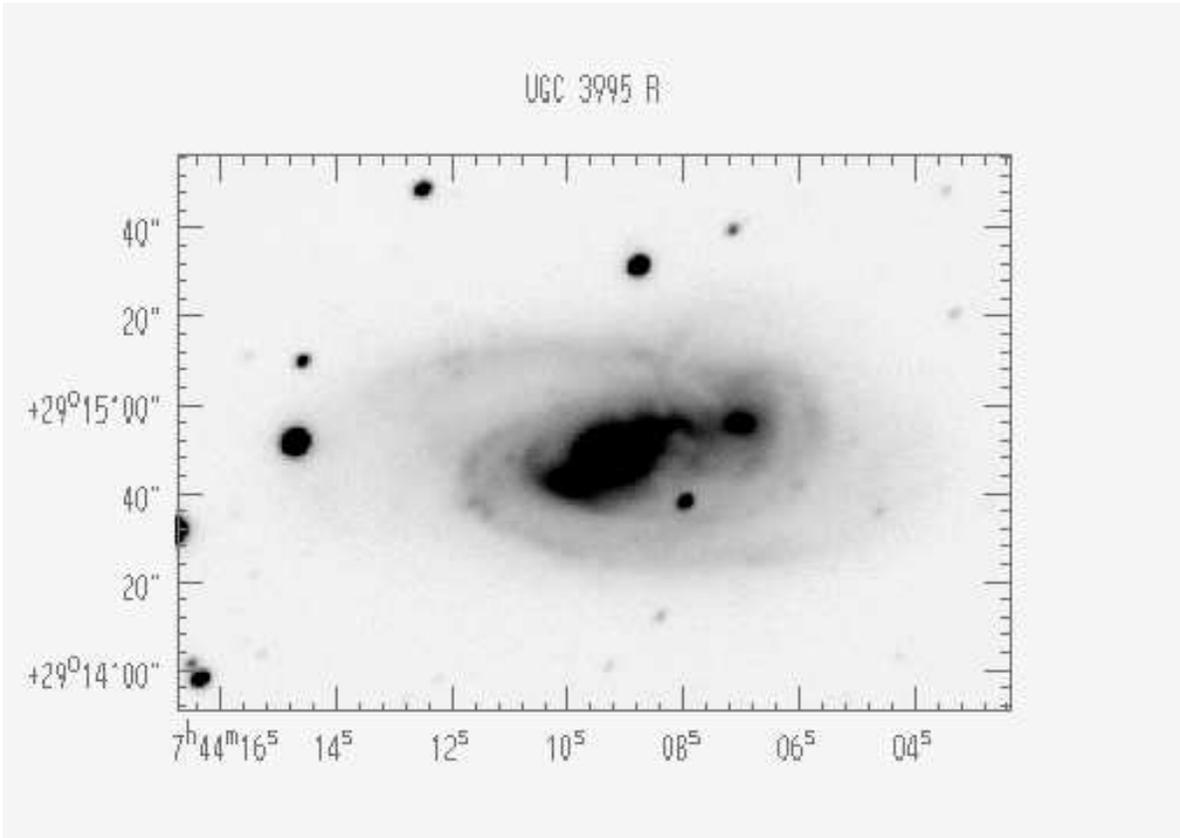} 
\caption[1]{Johnson R image of the UGC~3995 system (the average of three 
600$^{\rm s}$\ exposures. Note the perturbed appearance of the bar in the 
contact region between the two components, and the bright filament that seems 
to connect their nuclei. In this and the following labeled images, Right 
Ascension (abscissa scale) and Declination are refered to  J2000.  
\label{fig:R}} 
\end{figure}

UGC~3995 was selected as part of a larger investigation on the effects
of interaction on the gaseous  component in a small set of  galaxies
where activity might have been induced by mass-tranfer of tidally
stripped gas from a companion. UGC~3995 is a bright pair of spiral
galaxies with partially overlapping disks  and negligible redshift
difference that, at first glance, offers the possibility of direct
evidence in favor of tidal stripping and/or more quiescent cross
fueling into an active nucleus: a bright filament appears to connect
the Seyfert 2 nucleus of UGC~3995A with the nucleus of UGC~3995B over
$\approx$ 29.2 arcsec of angular distance, which corresponds to
$\approx$9 kpc of projected linear distance  (see Fig.  \ref{fig:R};
from the heliocentric radial velocity of UGC~3995A, $\rm v_r \approx
$4746 \kms, we compute a distance $\approx$ 63.3 Mpc; H$_0$ = 75 \kms
Mpc$^{-1}$\ is assumed thorough the paper). The earliest investigations
pointed out the peculiar nature of UGC 3995. UGC~3995 has been included
in the catalog of isolated pairs of galaxies (CPG 140, \cite{kara72})
where it is assigned the strongest interaction class LIN(br+ta)
intended to indicate the presence of tidal tails and bridges. The
``bridge'' is the bright filament seen in Fig. \ref{fig:R}. UGC 3995
was also listed as part of an isolated triplet, because a third galaxy
(MCG +05--19--003) of comparable brightness is located  $\approx$4
arcmin  NW (\cite{kara76}).  \cite{keel85} first identified the nucleus
of UGC~3995A as having Seyfert 2  properties; he suggested that the
UGC~3995 system was  ``a close pair of spirals one of which has a type
2 Seyfert nucleus and apparent knotty jet''. More recently, a report
based on integral field spectroscopy, and  high spatial resolution
(\cite{CVL98}) suggested that gas may be infalling toward the active
nucleus in the central regions of the UGC~3995  system.
 
From our data, unambiguous evidence for interaction and gas transfer
is difficult to find in the UGC3995 pair.   UGC 3995, in fact,
illustrates many of the  difficulties of unambiguously proving
interaction even for a very close pair of spirals with almost identical
redshift.  \S\ \ref{obs} summarizes our data and the reduction
procedures employed. \S\ \ref{results} presents the main observational  results of this
investigation. \S\ \ref{disc} considers UGC 3995 in the light of
generally accepted morphological, spectroscopic and photometric
indicators of interaction.  Preliminary results of this investigations
were presented in \cite{iau186}.
 
\section{Observations and Data Reduction \label{obs}} 

Narrow (redshifted \ha\ + \nii\ (filter width $\approx$ 70 \AA) and
broad-band (R, B) images of UGC~3995 were obtained on January 30, 1995
at the Cassegrain focus of the 2.1 m telescope of the Observatorio
Astronomico Nacional at San Pedro Martir (SPM), B. C., M\'exico.  A
Tektronix CCD (24  $\mu$m square pixel size, $1024 \times 1024$ format)
employed as a detector yielded a scale of 0.315 arcsecs/pixel and a
field of view of $\approx 5.4$ arcmin. Long slit spectra were collected
with the 2.1 m  SPM telescope on Dec. 9, 1994 (\hb\ spectral region)
and on February 2--4, 1995 (\ha\ region)  using a B\&Ch\ spectrograph,
with the same Tektronix detector. The December 1994 observations were
obtained with a 400 lines/mm grating and covered the usable range from
3900--6720 \AA, with a resolution $\approx$ 7 \AA\ FWHM.  A 1200
lines/mm grating was employed in January and February 1995 for
observations of the H$\alpha$\ region. The slit width was  200 $\mu$m
at the focal plane of the telescope (2.6 arcsecs). This setup  resulted
in a resolution of $\approx 2-2.2$ \AA\ FWHM. Table \ref{tab:obslog}
provides a log of observations, reporting the date and universal time
for each exposure, the exposure time, the filter or the spectral range
in case of spectra, and, for spectra,  the position angle of the slit.
A  500-s image obtained with the HST WFPC2 through the ``wide visual''
F606W filter on  May 17, 1994 was also retrieved from the HST data
archive (dataset {\tt U2E65T01T} see \cite{malkan98} for a full
presentation of that data). Part of the mosaicized image is shown in
Fig. \ref{fig:hst}.

\subsection{Data Calibration} 

\paragraph{Broad Band Images} Images  obtained with the same filter
were registered and co-averaged to improve S/N and to remove cosmic
rays.  Transformation from the instrumental to the Johnson photometric
system was achieved using a set of CCD standard stars in the globular
cluster NGC~4147 (\cite{ode92}). To compute the total luminosity of the
galaxy,  we  constructed growth curves by integrating the flux through
circular apertures, adopting the zero point and color terms derived
from the standard stars.
 
\paragraph{Narrow Band Images} Flux calibration of narrow band images
was obtained by observing a spectrophotometric standard star through
the narrow band filter.  After bias subtraction and flat field
correction,  the average R band-image was scaled and subtracted to the
average narrow band image  to eliminate the continuum underlying the
\ha+\nii\ blend. The resulting ``pure'' \ha\ + \nii\ image is shown in
Fig. \ref{fig:ha}. Uncertainties associated with the photometric
calibration and underlying continuum subtractions are estimated to be
within $\pm$ 20 \%.

\paragraph{Spectra} Wavelength calibration of science spectra was
obtained from HeAr comparison spectra,  taken between or after pairs of
30 minute exposures  with the same settings. The final wavelength zero
point was checked against the wavelength of strong sky lines
(\cite{marz95}).  An average shift of about $\rm \Delta v_r \approx$ 10
\kms\ was taken into account. Flux calibration was obtained by the
observation of tertiary standard stars, with two sequence of exposures
taken at widely different times during each night of the December and
January observing runs. Standard stars were observed with the slit
opened  to $\approx$ 500 $\mu$m in order to minimize light losses.
 
\section{Results \label{results}} 

\subsection{Broad Band Morphology \label{morph}} 
\begin{figure} 
\plotone{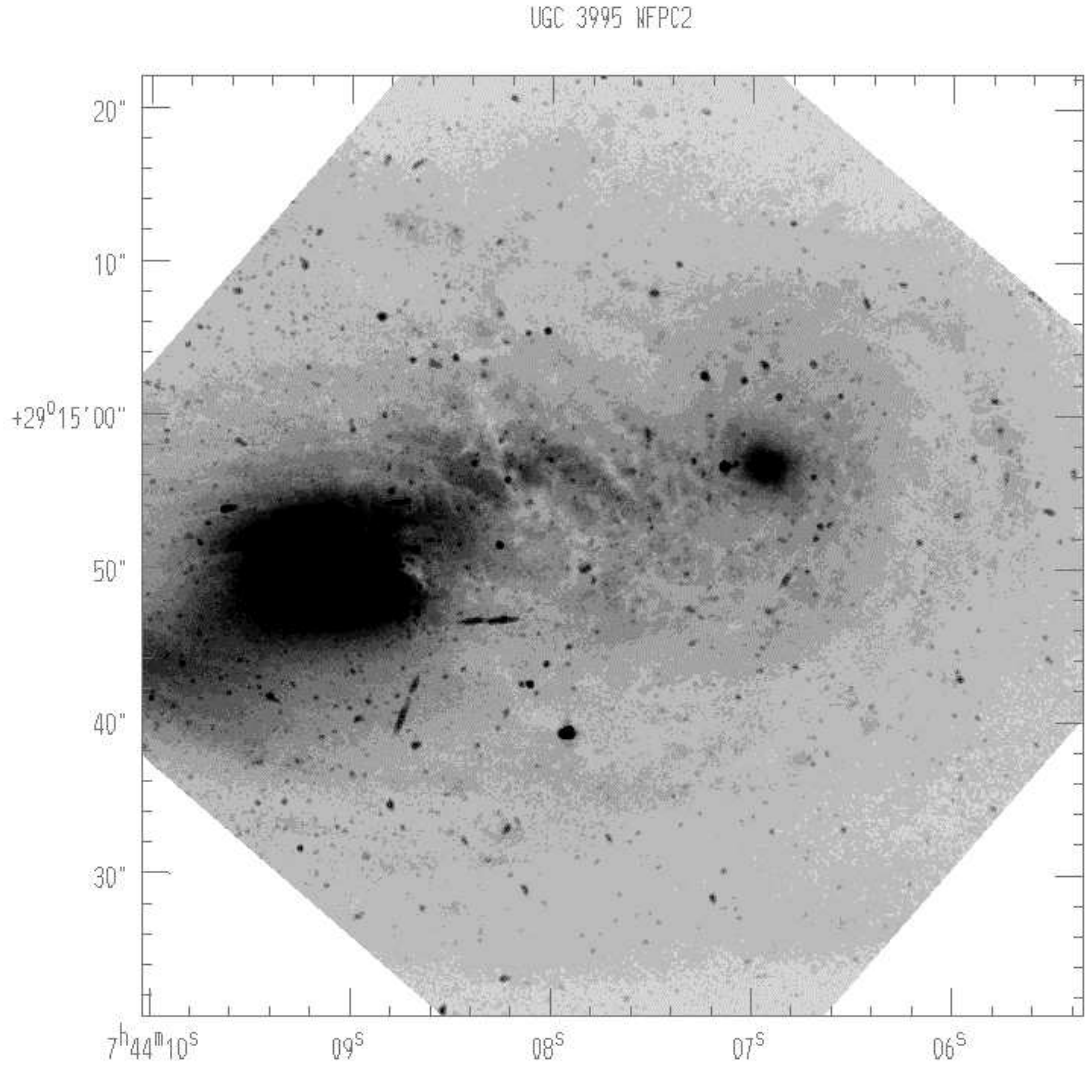} 
\caption[1]{WFPC2 image of the UGC 3995 system obtained through the wide visual 
F606W filter, after mosaicing and rotation by 49$^\circ$. Cosmic rays have  
been  partially removed following standard procedures. 
\label{fig:hst}} \end{figure}

Our R-band image (Fig.
\ref{fig:R}) suggests that the distorted appearance of the bright
filament joining UGC~3995A and UGC~3995B is almost entirely due to the
foreground spiral arms of UGC~3995B (and to dust lanes associated with
them), which cross the hammer-shaped bar like  sickle blades. The
eastern spiral arm of UGC 3995A obscures the bar of UGC3995A at its NW
end, creating the illusion of a filamentary region (the highest surface
brightness part of the bridge, which appear  ``semi-detached'' from the
inner part of the UGC 3995A bar) in between the nuclei of UGC 3995A and
B (clearly visible in both Fig. \ref{fig:R} and \ref{fig:hst}). The bar looks rather
symmetric, but some intrinsic asymmetry is associated with that
``semi-detached'' region, which appears to extend the  NW end of the
UGC 3995A bar  more than expected if the bar ends are symmetric with
respect to the nucleus.  The overall scenario is  validated,  with a
much greater extent of spatial details (including several arm branches
not visible in the ground based observations), by the WFPC2 image (Fig.
\ref{fig:hst}). In addition to the eastern arm of UGC 3995A, the
western arm of UGC 3995B can be followed as it winds around the galaxy
and also crosses UGC 3995A not  far from the nucleus, where a prominent
dust lane appears to cut across the circumnuclear regions.
 
\subsection{Broad Band Photometry \label{phot}} 

The first rows of Table \ref{tab:phot} report the integrated B, and R
magnitudes computed using the growth curve method,  with an expected
uncertainty of $\approx 0.1$ mag. Following  \cite{RC3}, if we  apply a
galactic extinction toward UGC~3995 of 0.09 magnitudes, at the adopted
distance of 63.3 Mpc  the absolute magnitude  of the whole system is
$\rm M_B  \approx -21.6$ mag.

\paragraph{Models of the Two Dimensional Surface Brightness Distribution 
\label{twodim}} In order to recover the individual photometric properties of
the two galaxies, we constructed a set of two dimensional models of their light
distribution. This was achieved by fitting only the regions of the frame
dominated by each galaxy separately. Regions including foreground stars and 
substructures such  as the bar, the knotted spiral arms,  the Seyfert nucleus 
and the nucleus of UGC 3995B (nuclear regions are dominated by seeing)  were
excluded from the fit by masking frames. We modeled the 2D  surface brightness
distribution of the galaxies with bulge and disk  components represented by
concentric ellipses whose apparent   flattening is due to  the inclination with
respect to the line of sight (see \cite{Byun}, \cite{d98} for references). The
bulge and the disk light distribution are given respectively by: 
 
\begin{eqnarray} \rm 
\mu_b &\!\!=\!\!& \rm \mu_e + \rm k \times \rm [(R_b/r_e)^{1/n} -1] \ \ \ \rm (bulge)\\ 
\rm \mu_d &\!\!=\!\!& \mu_0 +\rm  1.086 \times \rm (R_d/r_h)          \ \ \  \rm 
(disk) 
\end{eqnarray} 
 
where $\rm R_b = \{[(x_i)]^2 + [y_i/(b/a)_b]^2\}^{(1/2)}$ and $\rm R_d = 
\{[(x_i)]^2 + [y_i/(b/a)_d]^2\}^{(1/2)}$ are  the distances of the 
pixel $\rm (i,j)$ from the galaxy center. The light profile of the bulge 
follows a $\rm r^{1/n}$ law (\cite{CCD93}), while the disk has an exponential 
distribution.  The bar of UGC 3995A and the spiral arms have not been modeled. 
 
Table \ref{tab:phot} provides the results obtained for two ``best fitting'' 
models (yielding minimum $\chi^2$). Since the solution is not unique but
depends on the choice of the masking frame, our `fiducial solutions' are two 
models that give the best residuals,  a good behaviour of the global
growth-curve for two extreme, but equally reasonable choices of the masking
frames. Other choices of the masking frames are expected to give results
intermediate between model I and II. Finally, from the overall appearance and
the bulge/disk decomposition  performed, we suggest a morphological type SBbc
(or SB(r)bc) for UGC~3995A and Sc (or Sd) for UGC~3995B. 
 
\begin{figure} 
\plotone{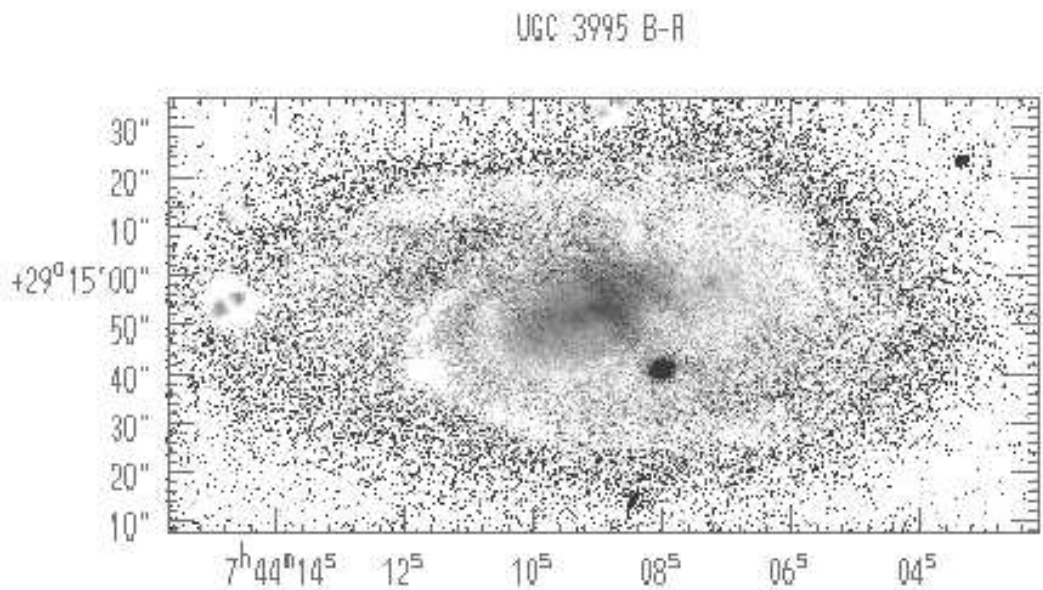} 
\caption[1]{ 
UGC~3995 color map $(B-R)$. The color range displayed is  approximately $\rm 
0.0 ~(white) \simlt \rm B-R \simlt 1.5 ~(black)$.    B-R 
values in correspondence of UGC 3995B are $\approx$ 0.7--0.8, 
suggesting that UGC~3995B is located in front of UGC~3995A. The  red spot
($\rm B-R \approx$ 2.0) south of the Seyfert nucleus is likely a  foreground star.
\label{fig:cmap}} 
\end{figure}

\paragraph{B--R Color Map}  The $\rm B-R$ color map (Fig. \ref{fig:cmap})  allows
us to trace the northern and southern spiral arms of UGC~3995A as  elongated blue
features ($\rm B-R \approx$ 0.5 -- 0.6). The bar appears redder  than the bulge
(average B--R, including the nuclear region, is $1.28\pm0.05$).  
UGC~3995B appears of rather bluer color index ($\rm B-R  \approx$ 0.7), which
suggests the absence of appreciable obscuration.    The  area in between the
nuclei is confused by the superposition of the NW end of  the bar in UGC 3995A
with the eastern arm of B. A comparison between the R and B--R images (and the HST
image) confirms that the spiral arms of UGC~3995B are  obscuring the western side 
of the UGC 3995A bar. The extension of the western  spiral arm of UGC~3995B is
most likely responsible for the reddish region that is elongated perpendicular to
the bar. 
 
\subsection{Emission Line Morphology \&\ Photometry} 
\begin{figure} 
\plotone{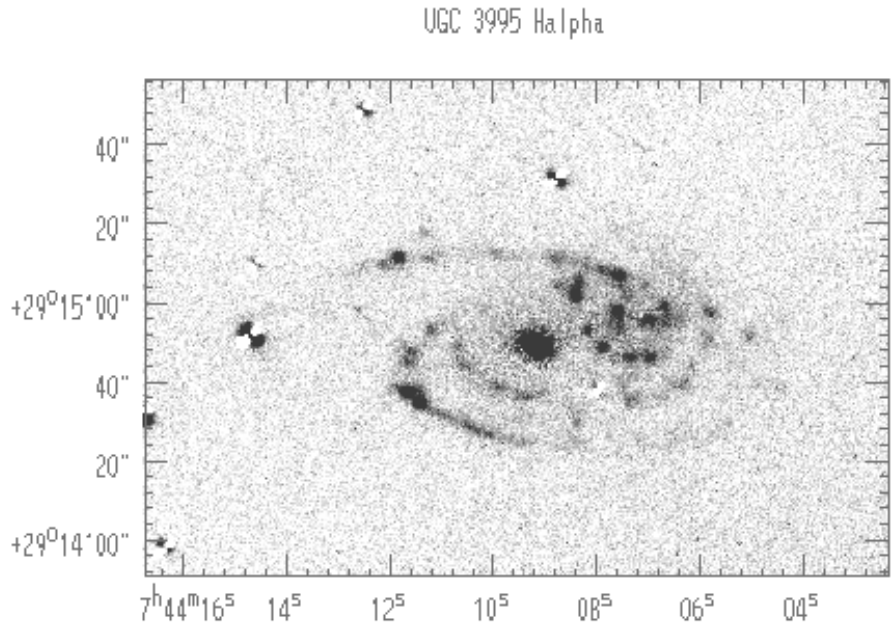} 
\caption[1]{\ha + \nii\ 
continuum-subtracted  image of UGC~3995. The image is the average of three 
1200$^{\rm s}$\ exposures. The continuum underlying \ha\ and \nii\ has been 
subtracted as a scaled R image. Strong stars show  bow tie residuals, probably 
because of slight difference in seeing or of small guiding errors at the time the narrow and R band images 
were collected.  \label{fig:ha}} 
\end{figure}

\begin{figure} 
\plotone{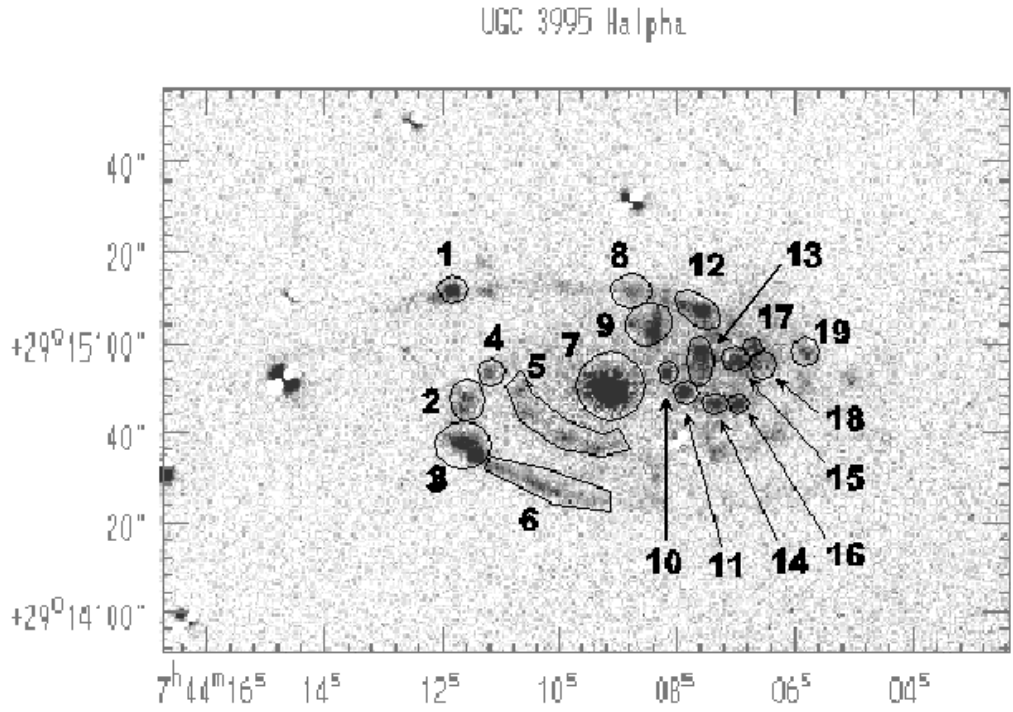} 
\caption[1]{\ha + \nii\ 
continuum-subtracted  image of UGC~3995, same as Fig. \ref{fig:ha}.  
Labels have been assigned in order of decreasing rights ascension. Fluxes and a 
short description of each emitting region are reported in Table \ref{tab:ha} 
  \label{fig:halabeled}} 
\end{figure}

The broad band morphology can be easily compared to the morphology  of the ionized gas, which results from the \ha\ + \nii\ map shown in Fig. \ref{fig:ha}. The prominent bulge and the bar of UGC 3995A almost completely disappears in the  narrow band image, where most 
emission can be ascribed to  (i) the galactic nuclei, (ii)  knots associated to 
the spiral arms of either galaxies, (iii) weak  \ha\ diffuse emission. The 
absence of strong unresolved, diffuse \ha\ emission is confirmed by our long 
slit spectra at P.A. =45$^\circ$, and 125$^\circ$: \ha\ emission on the eastern side of   UGC~3995A is, for example, 
detected only in correspondence of spiral arms or knots. Table \ref{tab:ha} 
lists region  label, position with respect to  UGC~3995A,  and fluxes.    Fig. 
\ref{fig:halabeled} identifies each emitting region with a number in rough 
order of decreasing right ascension.  Diffuse emission 
is weak but  not negligible, since emission that cannot be ascribed to discrete 
sources listed in Tab. \ref{tab:ha} appear to contribute  to $\simlt$ 1/2 of 
the total \ha +\nii\ emission. The diffuse emission is subject to an 
uncertainty  which is estimated to be $\simlt$ 50 \%, larger than that of the 
flux of discrete sources ($\approx 20$ \%). 
 
The knots on the eastern spiral arm of UGC~3995B, and several knots of the 
northern arm of UGC~3995A create the appearance of a ring, which is  an optical 
illusion due to a chance peculiar orientation of the two spiral arms. It is 
important to note, in this respect, that the knot at the center of this 
pseudo-ring (labeled 13) is not the nucleus of UGC~3995B, which is  knot 15, 
located at the center of symmetry of the spiral arms of UGC~3995B. This is seen 
clearly comparing Fig. \ref{fig:ha} and Fig. \ref{fig:R}, which have the same 
scale.  The knot 13 at the center of the pseudo-ring cannot be attributed to any 
of the spiral patterns of the two galaxies. This region  is located at the end 
of  the  high surface brightness extension of the UGC 3995A bar NW tip.   The main
features emerging from the previous analysis are sketched in Fig.
\ref{fig:sketch}.  

\subsection{Nuclear Spectroscopy} 

\begin{figure} 
\plotone{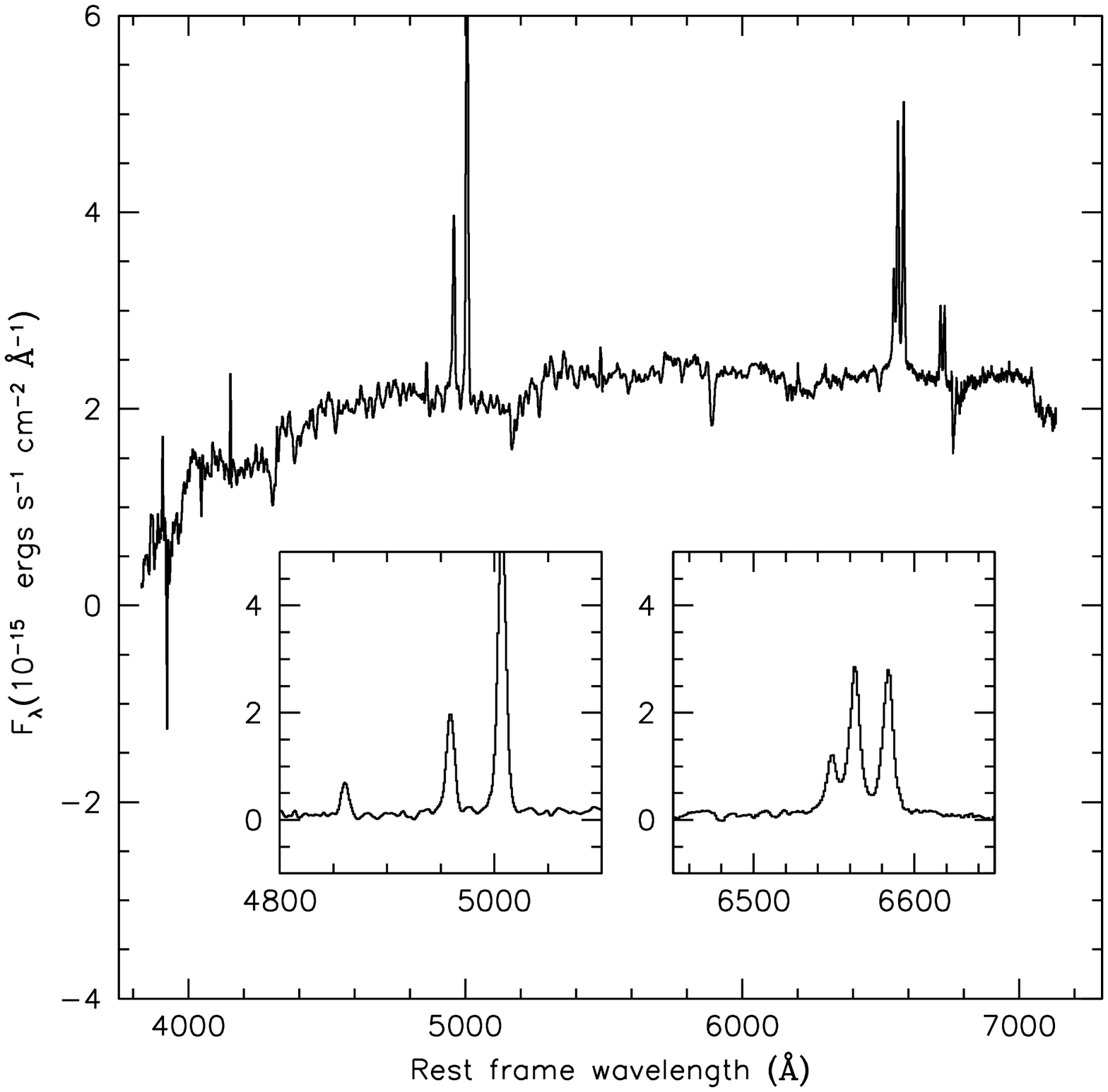} 
\caption[1]{De-redshifted nuclear spectrum of UGC~3995. Abscissa scale is 
wavelength, ordinate scale is specific flux multiplied by 10$^{15}$. The two 
smaller inlets show the \hb\ and \oiii\ region (left) and the \ha\ + \nii\ 
blend (right) after subtraction of the galaxy absorption spectrum, modeled
employing a scaled spectrum of
NGC 3379, obtained from Kennicutt's digital spectrophotometric atlas of
galaxies (\cite{ken92}).  \label{fig:nucspec}} 
\end{figure}

\paragraph{UGC 3995A} The nucleus of component A shows a type 2 Seyfert
spectrum
(\cite{keel85}). Fig. \ref{fig:nucspec}  shows the flux calibrated nuclear
spectrum (in the rest frame of the galaxy, not corrected for galactic extinction)
from  3800 to 7200 \AA.  Measurement of the lines in the  UGC~3995A are reported
in Table \ref{tab:eml}.   The radial velocity, measured from \ha, \nii, \sii, and
\hb\ and \oiii\ lines in all our best spectra is $\rm v_{r,hel} = 4746 \pm 18$
\kms, where the uncertainty is at 2 $\sigma$ confidence level. This determination
agrees  with the value  $\rm v_{r,hel}\approx$ 4747 \kms\ obtained from the \hi\
21-cm profile (\cite{sul83}; see Fig. \ref{fig:rv}). The spectral energy
distribution shown in Fig. \ref{fig:nucspec} suggests an overall spectral type of
G2 (EW(\hb) in absorption  $\approx$ 1.59 \AA, summed over $\approx$ 7'').

\paragraph{UGC 3995B} In Table \ref{tab:eml} only the fluxes for \ha, \niis, and
\sii\ are reported.  A spectrum at P.A. =103$^\circ$\ kindly provided us by W. C.
Keel shows \oiiis/\hb $\approx$ 0.20$\pm 0.05$  confirming that the spectrum is of
a low-ionization nuclear \hii\ region. The radial velocity of the single, middle
peak in the 21 cm \hi\ profile ascribed to UGC~3995B (see next paragraph, \S
\ref{extended}) is $\rm v_{r,hel} \approx 4748$ \kms. The spectra  obtained at
P.A. = 106$^\circ$ yield a radial velocity $\rm v_{r,hel} \approx 4766 \pm 16$
\kms,  exceeding by 27 \kms\ that of UGC~3995A (the excess is much larger than the
internal consistency errors of our spectra), in fair agreement with the \hi\
determination.
 
\subsection{Spectroscopy of Extended Emitting Regions \label{extended}} 
\begin{figure} 
\plotone{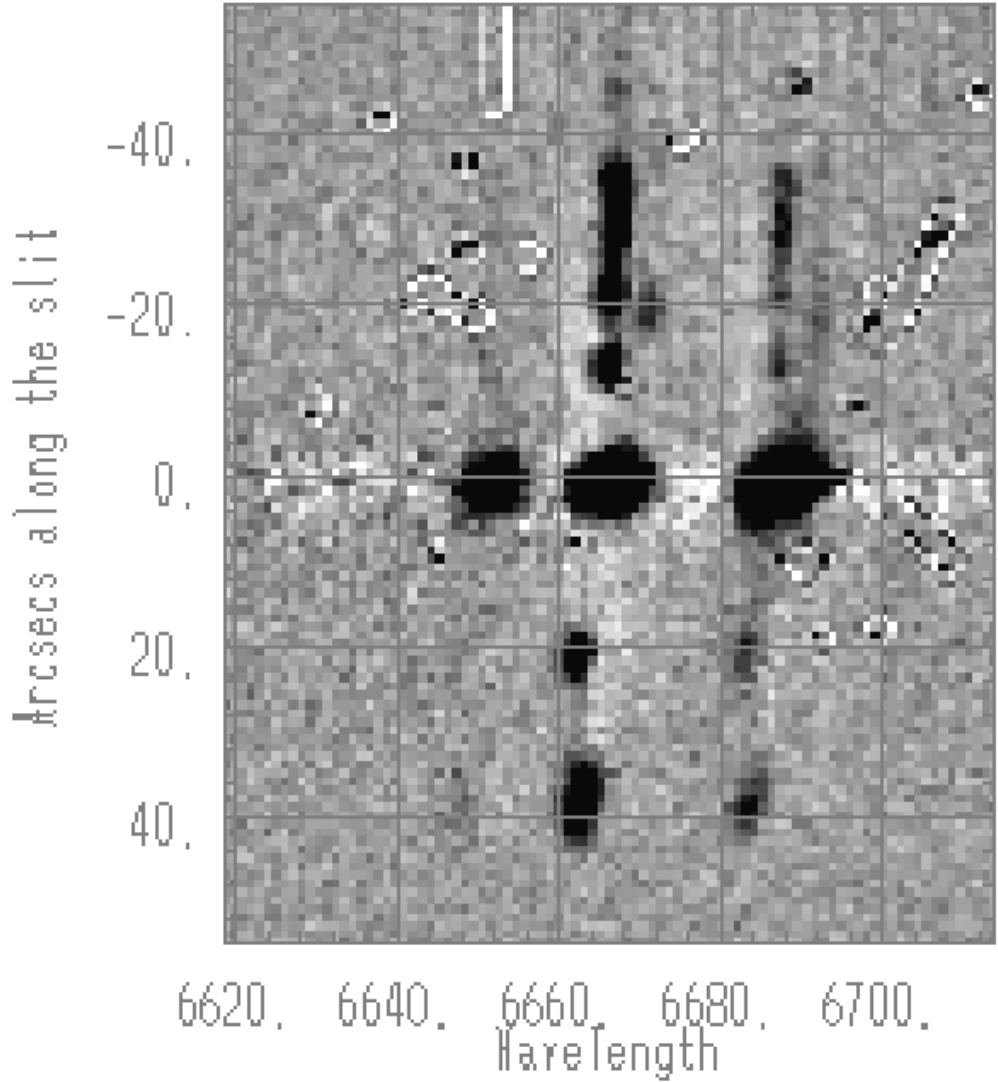} 
\caption[1]{Long slit 
spectrum at P. A. = 106$^\circ$, with both galaxy nuclei in the slit. 
Abscissa is wavelength in \AA; ordinate is  arcsecs along the slit. The 
zero-point of the angular scale has been set at the position of the Seyfert 
nucleus. The spectrum is the average of four 1800$^{\rm s}$\ exposures 
obtained sequentially, cleaned for cosmic rays.  The scale of
intensity has been set to show the important details  at a level which
is just $\approx$6$\sigma$\ the noise level.  The  galaxy stellar and
the AGN continuum have been removed employing a median filter to better
show the structure of the extended emitting regions. \label{fig:pa106}} 
\end{figure}

\begin{figure} 
\plotone{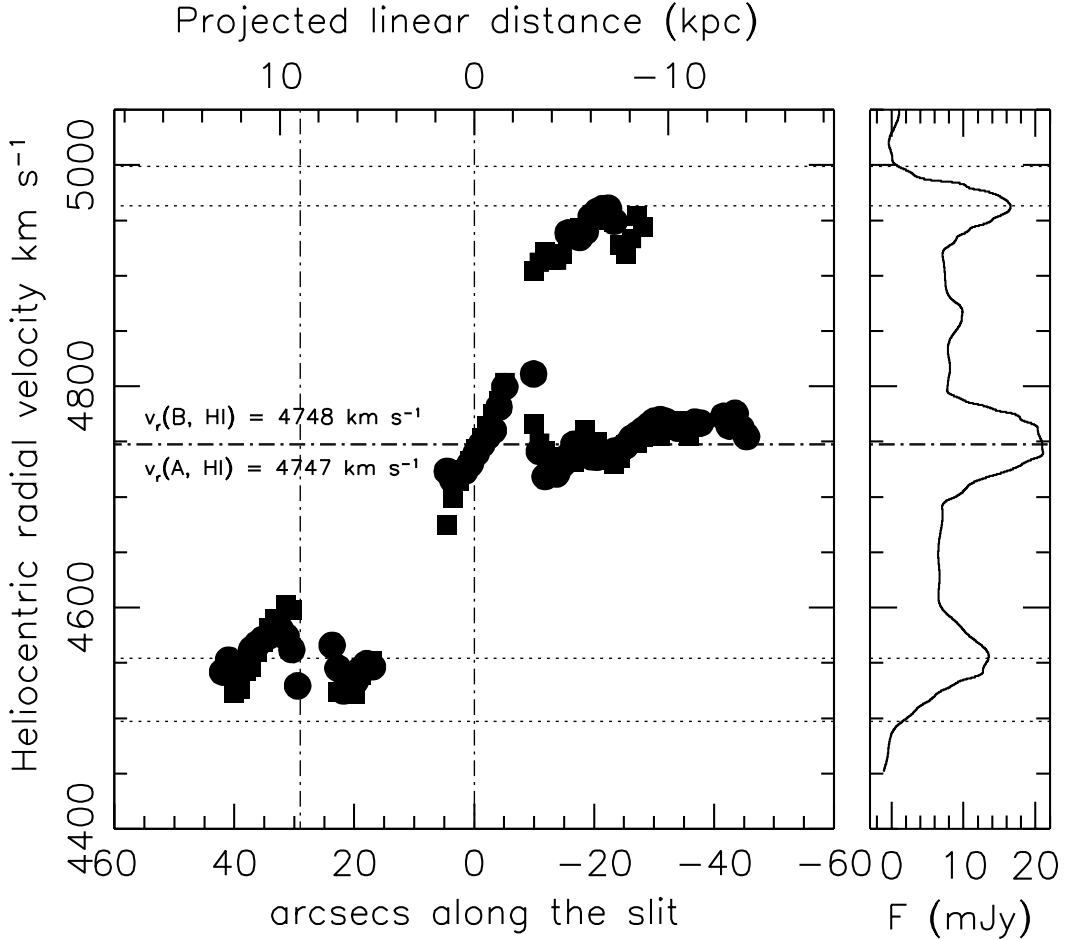} 
\caption[1]{Left Panel -- Radial velocity curve at P. A. = 106$^\circ$. The 
abscissa scale  is arcsecs (lower axis) and projected linear distance (upper 
axis) along the slit increasing to the West  (the zero  point of the angular
and spatial scale has been set at the position of the  Seyfert nucleus), the
ordinate  scale is  heliocentric radial velocity in \kms.  Filled circles:
\ha\ measurements; filled squares: \niis\ measurements. Uncertainties
associated with measurements are estimated to be, at a 2$\sigma$\ confidence
level, $\pm 50$ \kms\   for the obscured UGC 3995A emission at $\rm v_r
\approx 4950$ \kms, $\pm$ 10 \kms\ for measurements in the nuclear region and
of the UGC 3995B \ha\ line, and approximately $\pm$ 30 \kms\ in all other
cases. The horizontal dotted-dashed lines are the heliocentric radial
velocities for UGC~3995A and 3995B, from the 21 cm \hi\ profile of
\cite{sul83}. The  dotted lines are the  heliocentric radial velocities of
the ``horn'' peaks and  of the 20 \%\  intensity level in the 21 cm line
profile.   In correspondence of  UGC~3995B, gas from both UGC~3995A and
UGC~3995B is detected. The lower radial  velocity  part of the velocity curve
is from gas associated with UGC~3995B.  \label{fig:rv} Right Panel -- \hi\ 21
cm line profile; ordinate is radial  velocity as for the right panel;
abscissa is specific flux in units of milli Jansky. } \end{figure}

Four
co-averaged 30 min. exposures yielded the long slit extended \ha\ + \nii\ spectrum
shown in Fig. \ref{fig:pa106}.  On the western side of UGC 3995A (upper part of
Fig.  \ref{fig:pa106}) the strongest emission is due to gas in the disk
of UGC~3995B. Emission from UGC 3995A is very weak, because of obscuration
by the foreground disk of UGC 3995B. Only the  \ha\ knot on the higher velocity
side (region 13)  and the very faint and extended \niis\ emission can be
associated to UGC~3995A. This interpretation is made more obvious by the radial
velocity curve at P. A. $\approx$ 106$^\circ$ (left panel of  Fig.  \ref{fig:rv}),
obtained from  measurements on the spectrum shown in Fig.  \ref{fig:pa106}.    The
nearly flat radial velocity curve to the West  is due to the brighter \ha\ and
\nii\ emission in Fig.  \ref{fig:pa106}, and is consistent with a galaxy observed
nearly face-on,as it is the case of UGC 3995B.  The inclination of UGC~3995B can
be estimated from the slope of the radial velocity curve. Assuming an average
slope  for Sc galaxies of 37 \kms\ kpc$^{-1}$\ in the inner part of the rotation
curve (\cite{rubin80}), we obtain i(UGC 3995B) = 8$^\circ \pm 2^\circ$. The high
velocity part of the radial velocity curve is due to the western side of the
UGC~3995A disk which is heavily obscured. Consistently, its radial velocity is
close to that of the high velocity horn in the \hi\ 21 cm line profile (right
panel of Fig.  \ref{fig:rv}, from \cite{sul83}).

It is  noteworthy that for UGC 3995B we observe both \ha\ and \nii, with the 
first stronger, as expected from normal disk \hii\ regions (the  \hii\ nature of
the UGC 3995B knots across the slit at P. A. =106$^\circ$\ is confirmed by a
spectrum provided us by W. C. Keel, which  shows \oiiis/\hb $\simlt$ 0.5 for the
extended emission). On the contrary, for UGC 3995A we  observe (save in region 13)
\niis\ but no \ha, which most likely falls below  $3 \sigma$\ the noise level.
This implies that the ratio \niis/\ha\ is $\gg$ 1, possibly  $\simgt$ 2.5. 
Appreciable \niis\ emission without \ha\ is observed on  the eastern side of UGC
3995A as well, up to $\approx$ 10 arcsec from the nucleus (see Fig.
\ref{fig:pa106}). A possible cause may be related to an ionization cone 
(\cite{CVL98}). It is unlikely that extended \niis\ emission is due exclusively 
to under/over subtraction of sky continuum/lines or other instrumental  effects,
especially on the western side of UGC 3995B. 
There is also  some evidence for an \niis\ emitting  filament on the western side of
UGC3995A (barely visible in Fig. \ref{fig:pa106}) within 10 arcsecs from the
nucleus. This emission appears to join the high velocity end of UGC3995A
nuclear emission with the \niis\ emission associated with UGC 3995B (it gives
rises to the ``curl'' at the eastern end of the UGC 3995B radial velocity
curve, Fig. \ref{fig:rv}). The  weakness of the high velocity \niis\ extended
emission makes it  desirable that confirmatory data be obtained.  

\subsection{The Optical Thickness of UGC~3995B \label{thick}} 
 
\begin{figure}
\plotone{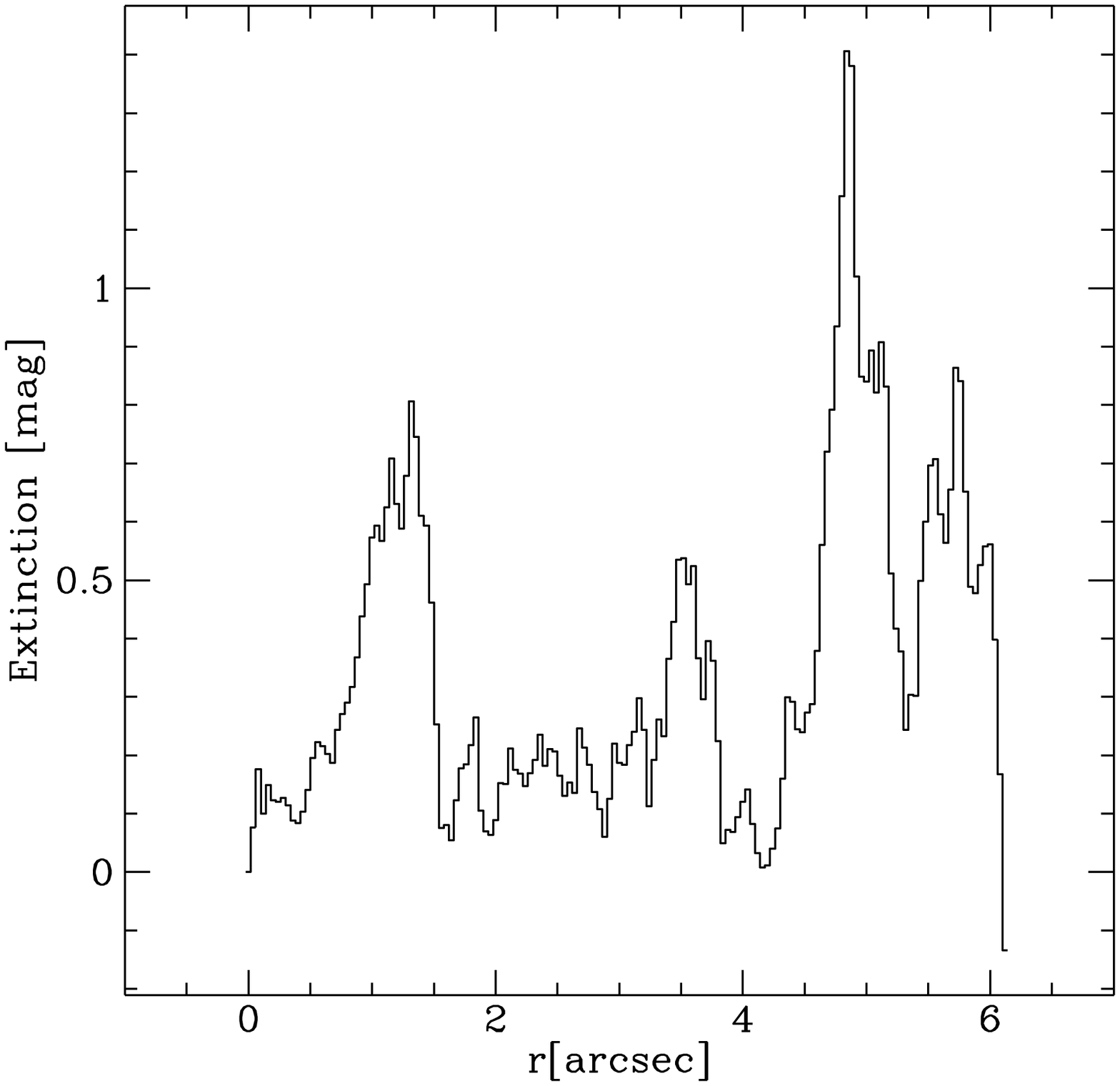}
\caption[1]{Extinction profile along the bar of UGC 3995, measured on the
WFPC2 image, in wide-V magnitudes (2.5$\rm
\log(I_{unobscured}/I_{obscured}$)). The origin of the abscissa scale has
been set at the Seyfert nucleus of UGC 3995A. Angular distance d'' is
increasing toward the NW; at d'' $\approx$  1  the absorption peak is
associated with the eastern extension of the western spiral; at d''$\approx$
5 the peak of absorption is due to the eastern spiral arm.
\label{fig:lastplot}} \end{figure}

The UGC~3995 system is one of the rare cases in which one galaxy appears in 
foreground of another, without any gross morphological disturbance. 
The optical thickness of
galaxy disks has been a hot topic in recent years  (\cite{vale90}, \cite{wk92},
\cite{puzz93}, \cite{domin98}), since  recent results suggest  significant optical
depth, at variance with early studies (\cite{holm58},  \cite{deV59}).  \cite{wk92}
have argued that the optical thickness is highly variable across a spiral disk and
our images (especially WFPC2) of UGC 3995 support this view.  The problem cannot
be extreme, as the northern spiral arm of  UGC~3995A can be traced in the region
behind UGC~3995B.  An estimate of the  absorption $\rm \langle A_V \rangle$ of 
the Sc galaxy UGC~3995B can  be obtained directly from the broad band WFPC2 image.  If the
UGC~3995A bar is roughly symmetric across the nucleus and unperturbed (which seems
to be the case save at its NW end), then the galaxy surface brightness measured
along the bar axis should be the same in both directions.  We subtracted a model
of component B to the WFPC2 image; we then  extracted the photometric profile {\em
along the bar} across the Seyfert nucleus. The photometric profile on the obscured
side was then divided  by the folded photometric profile of the unobscured side.
The results is shown  in Fig. \ref{fig:lastplot}. Absorption in the inter-arm
region between the extension of the western arm and the eastern arm is fairly
constant (UGC3995A appears to be dimmed as if seen through a screen of absorbing
matter), the median $\rm \langle A_V \rangle$ being  $\approx 0.18$, with
intrinsic scatter $\pm$ 0.09 mag (2$\sigma$), in good agreement with the values
found for other spirals (\cite{domin98}). $\rm A_V$\ is much higher in
correspondence of the spiral arms: 0.9 mag.  and 1.4 mag for the western
extension
and for the eastern arm respectively (at $\approx$ 1 arcsec and 5 arcsec in Fig.
\ref{fig:lastplot}). Absorption is quite patchy in both the interarm region and in
the spiral arms. In correspondence  of the darkest patches on the dust lanes, $\rm
A_V$ is estimated to be in the range 2--4 magnitudes.  

We considered also two approaches based on our spectra at P.A.=106$^\circ$: 
(a) we assume that the integrated  \niis\ intensity is intrinsically the same
on the eastern (unobscured)  and western (obscured) side of UGC~3995A; we
obtain I(\niis)$\rm  _{East}$/I(\niis)$\rm  _{West} \approx 0.37$, which
corresponds to $\rm A_V  \approx 1.33$; (b) if we do the same assumption for  
the \ha\ integrated  emission we obtain I(\ha)$\rm _{East}$/I(\ha)$\rm _{West}
\approx 0.31$, thus  $\rm  A_V$\ results  $\approx 1.59$. The average of these
values is $\rm  \langle A_V \rangle \approx  1.46$, in good agreement with 
estimates for other spirals from the Balmer decrement (\cite{puzz93}), and in agreement with
our estimate for the spiral arms. Due to the  patchiness
of absorption, and to selection effects which favor the brightest \hii\
regions, the  value obtained from the spectra should be considered as an  upper
limit.  Additional caution  is warranted  by the
possible variation of  ionization  conditions across the disk of UGC~3995A (\S
\ref{extended}).

\section{Discussion \label{disc}} 
  
\subsection{Absence of Strong Tidal Effects} 

There is little evidence in  the broad-band ground-based or HST images for
features that can be  unambiguously assigned to the effects of a strong tidal
field. No  evidence of {\em connections}  is seen in the  narrow band
image (Fig.  \ref{fig:ha}). Taking into account the confusion due to overlap 
of the two galaxies, the spiral arms of  UGC~3995B  appear nearly symmetric
with respect to the nucleus and with no evident  sign of perturbation. 
 
The \hi\ velocity profile  is remarkably  regular, if compared to
several  interacting galaxies also studied  by \cite{sul83}. A reasonably
straightforward deconvolution of that profile  is possible because UGC 3995A and
B  show considerably  different inclinations to the line of sight (Fig.
\ref{fig:rv}).  Component A shows a ``double-horn'' profile characteristic of a
fairly  massive inclined spiral galaxy. A third horn is detected near the  center
of the broad ($\rm \Delta v_r = 476$ \kms) double horn that is most  simply
ascribed to the single-peaked emission from the near face on  component B.  The
sides of the broader profile are quite steep and  characteristic of an
undisturbed  \hi\ disk in a spiral galaxy.  \cite{sul83} present a qualitative
classification of \hi\ profiles in  terms  of evidence for gravitational
disturbance.  In that scheme  UGC3995 is normal except for the presence of the
third peak. If the  \hi\ disk is  sensitive to dynamical perturbations from near 
neighbors,  interaction has not yet produced any appreciable  disturbance of the
\hi\ profile.  Simulations show that disk \hi\ gas  is the most component most
sensitive to tidal forces (e. g.  \cite{combes97}, and references therein). 

The total  line luminosity for the UGC 3995 system  measured from the continuum
subtracted narrow band image (Fig. \ref{fig:ha}) is L(\ha +\nii) = $2.8\times
10^{41}\rm ~ergs~ s^{-1}$. If we subtract the contribution from the Seyfert
nucleus, and assume a ratio \nii/\ha $\approx$ 0.5, $\rm L(H\alpha) = 1.5\times
10^{41}  ~ergs ~s^{-1}$. This corresponds to a total star formation rate SFR 
between $\rm 0.1 - 100~ M_\odot$ $\approx$ 1 $\rm ~M_\odot yr^{-1}$\ which is far
below the SFR of classical starburst galaxies. The luminosity  of the brightest
emitting regions in both galaxies are not extraordinary for late-type spirals  (e.
g., \cite{kenn89}). The \hb\ absorption EW  value is also consistent with the
absence of a young stellar population (\cite{bressan96}). 

\subsection{Evidence of Weak Tidal Disturbances\label{int}} 
  
The components of UGC3995 are proximate on the sky and in redshift but  our
images fail to confirm any obvious {\em strong} tidal feature, as well as
enhanced star formation.  We do find  several second-order features that
suggests that the two galaxies are  not double simply because of a chance
superposition  (as in the case of NGC 3314; \cite{schw85} where two galaxies
of different redshift appear overlapping,  or of a widely separated pair at
roughly the same distance from the Galaxy),  but that they are actually
proximate in space  and physically interacting:  (1) emission knots in the
spiral arm of UGC 3995B on the  side towards A are much brighter than those on
the opposite side; (2)  emission knot 13 (see Fig.  \ref{fig:ha}); (3) the
appearance of the northern spiral arm of UGC 3995A; (4) the ``semi-detached''
region at the NW end of the UGC3995A bar; (5) a possible extension and
distortion of the western spiral arm of UGC3995B, which is appreciable
especially in the  HST image (Fig. \ref{fig:hst}).
 
Point (1) is the strongest evidence in favor of appreciable tidal effects. 
\ha\ images are somewhat sparse in the literature, and most intensity
information is lost in print, so that a systematic and accurate comparison 
between UGC 3995 and other systems is not possible. Inspection of published
\ha\ images (for examples those by \cite{hodge75a}; \cite{hodge75b}; 
\cite{hodge83}, \cite{ryder93}) suggest that the distribution of \hii\ regions
in isolated late-type spirals are usually rather symmetric. Contrarily bright
arc-like features are found in several interacting objects (e.g. UGC 4081,
\cite{CVL98}; NGC 7469, \cite{marquez94}, \cite{pron90}), although not always
on the side towards a companion.  Surveys of the distribution of \hii\
regions  confirm that bridges and arcs of prominent \ha\ emission are more
frequent in peculiar and interacting galaxies. The \ha\ asymmetry in the arms
of UGC 3995B is {\em very strong} and similar to the kind seen in the cited
interacting systems. The \ha\ + \nii\ intensity ratio between the eastern and
western disk emission in UGC3995B is  $\simgt$ 2. An asymmetry  of this
amplitude is rarely found in  published \ha\ images of isolated galaxies,
although the distribution  and intensity of \ha\ emission is  never perfectly
symmetric in  late type spirals. The enhancement is consistent with the idea
that the two galaxies are  approaching, primarily, on plane of sky -- with the
near side gas showing the first evidence of an interaction enhancement. 
 
\begin{figure} 
\plotone{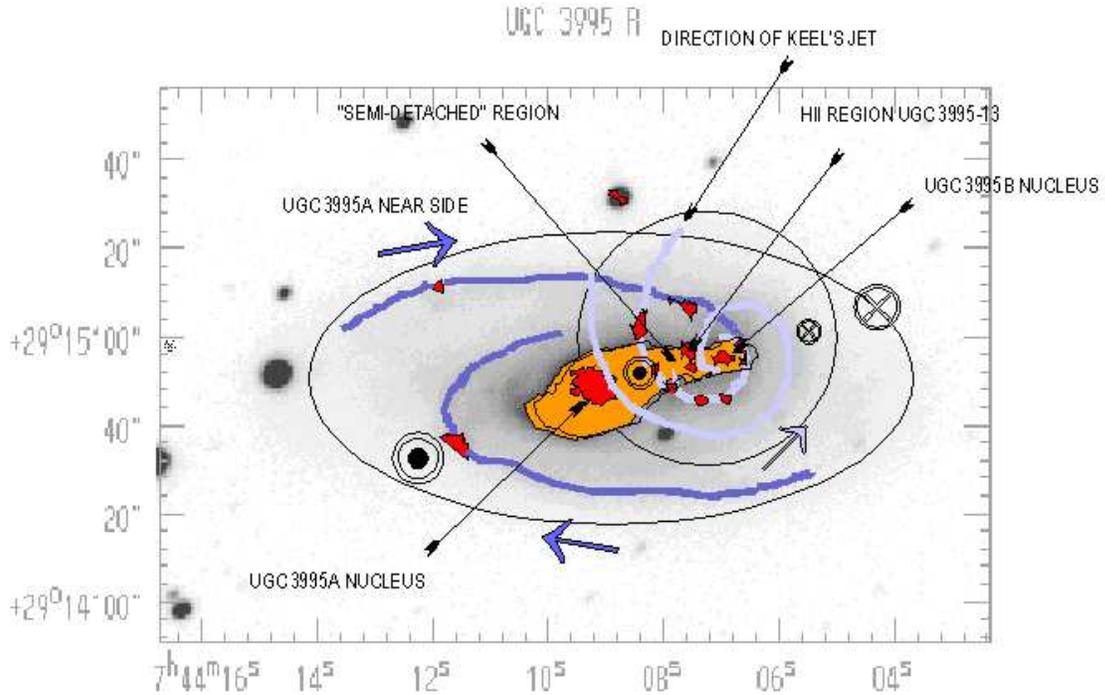} 
\caption[1]{Graphical sketch of the UGC 3995 system. Orange color: bar and 
filament connecting the nuclei of UGC 3995A and B, the highest surface 
brightness region of the system. Dark blue: spiral arms of UGC 3995A; light 
blue: spiral arms of UGC 3995B. Red: emitting regions as in \ha\ + \nii\
map;  The $\bigodot$\  marks negative radial velocity; the $\bigotimes$\ 
symbol positive radial velocity. The large symbols are for UGC 3995A. The
``jet''  (\cite{keel85}) appears to be  associated with a spiral arm of
UGC~3995B that extends toward objects which are, most likely,  a foreground
star, and a very faint background galaxy (\cite{CVL98}).}    
\label{fig:sketch} 
\end{figure}

2) Region UGC 3995-13 in the \ha\ + \nii\ map is located between the nucleus 
of B  and the bright \ha\ knots discussed in point 1. It cannot be
unambiguously assigned to the spiral arms of either galaxy.  The radial
velocity demonstrates that knot 13 belongs to UGC 3995A and is seen through the
disk of B. The ratio \niis/\ha $\approx$ 0.4 suggests that it is an \hii\
region located at the end of a ``semi-detached'' high surface brightness region
(see Fig. \ref{fig:sketch}). It may be therefore part of the ``bridge.'' Knot
13 is in any case difficult to explain if UGC3995A and B is a chance
projection.  

3) The northern arm of UGC 3995A  has a more diffuse appearance, suggesting the
possibility of a perturbation.  This is appreciable especially in the POSS
plates and in the B image after subtraction of the two dimensional galaxy
models (not shown).
 
4) The bright filament motivated the ``bridge'' classification in the CPG. As
a matter of fact, the bridge appears to be due to: (a) the NW side of the UGC
3995A bar; (b) the ``semi detached'' high surface brightness region, which
includes the NW end of the hammer shaped bar isolated by the eastern arm of
UGC 3995A; (c) a region that includes UGC 3995-13, which appears to join the
other features to the nucleus of UGC 3995B. A model based on reflection of the
unobscured end of the UGC 3995A bar, coupled with the overlapping spiral arms
of UGC 3995B,  does not entirely explain the extension of the
``semi-detached'' feature, which may have been tidally deformed.  However, the
nature of region (c) is not free of ambiguity because the surface brightness
measured on the HST image appears to be only marginally higher ($\simlt$ 20\%) than its symmetric counterpart on the western side of UGC 3995B. On the R
image, which is contaminated by \ha\ emission, the surface brightness
difference is more evident. 

5) As noted in \S\ \ref{morph}, the western arm of UGC 3995B appears 
to extend much farther than its symmetric counterpart. The two arms are 
reflections of one another for the first 180$^{\circ}$ of winding.  The 
western arm extends towards component A and crosses the circumnuclear 
region of that galaxy. It is visible as a prominent dust lane, 
especially, on the HST image. It is tempting to ascribe the extension 
as evidence of interaction and the analogy to the extension of an arm 
of M51 towards its companion is obvious (e. g., \cite{fins98}). 
On the other hand, the apparent lack of symmetry for the eastern and
western spiral arm of UGC 3995A  may be due only to the lack of background 
emission on the western side of UGC 3995B: the eastern
extension of the  western arm can be noted only because it adsorbs light from
UGC 3995A. 

Concluding, the most robust evidence of interaction is due to the brighter
\ha\ emission on the eastern side of UGC 3995B, and to region UGC 3995-13,
which is right at the end of the high-surface  brightness distortion of the
bar (see Fig. \ref{fig:sketch}). Other features (points (4) and (5) above) are
more ambiguous in their interpretation.  
 
\subsection{The Encounter} 

The relative orientation of UGC 3995A and B is well  constrained. The Sc
galaxy  UGC~3995B is oriented almost face-on ($\rm i  \approx$ 8$^\circ$) and
is located in front of the SBbc galaxy  UGC~3995A ($\rm i  \approx$ 60$^\circ$).  Component B  partially  obscures the western side of component
A.  If spiral arms trail then  the northern side of UGC ~3995A is the near
side. A similar consideration  suggests that the southern side of UGC 3995B is
nearer. However, the line of  nodes for UGC 3995B is unconstrained. 

A dominant velocity component along the E $\rightarrow$ W direction would
imply a prograde encounter.  The system does not show  any of the
morphological features expected from such a prograde encounter {\em after} closest
approach. The global symmetry of both galaxies is more consistent with an
encounter stage  before perigalacticon. In this case, the western arm of UGC 3995B
could have been really distorted towards UGC 3995A as the two galaxies should have
already passed in front of each other on the plane of the sky (see point (5)
of \S \ref{int}).  The observed
\hi\ 21 cm profile argues against the prograde encounter scenario, as any
velocity perturbation of disk gas (not to be confused with the orbital
velocity of the two galaxies) would have to be entirely on the plane of the
sky.

The alternative possibility involves a velocity component along the 
direction W $\rightarrow$ E. In this case, the encounter would be 
retrograde, with the two  galaxies either before or after their closest 
approach.  Among interacting Seyfert galaxies, a significant fraction appears
to undergo a retrograde encounter (\cite{keel96}). A retrograde encounter may
be as efficient as a prograde one in driving gas from the inner central
parsecs of a gas rich galaxy like UGC 3995A toward the central black hole.  
A retrograde encounter is consistent with the unperturbed  appearance of  the
\hi\ disk, and with the presence of localized star  formation in knots along
the spiral arms. A retrograde encounter,  always assuming that the galaxies
are at, or close to perigalacticon,  would not produce large-scale
distortions. Test particles in a  restricted three body simulation are
subject  to forces pulling them  alternatively inward and outward, with
little net result  (\cite{binney}, \cite{toom72}). These kind of forces are
expected to  enhance the probability of molecular cloud collisions, and hence
to  favor localized star formation especially in spiral arms, as observed  in
UGC 3995.  
 
The bar appears strong, regular, and dominated by stellar emission.   A tidal
perturbation due to the proximity of a companion may be enough to induce the
formation of such a bar, which is though to facilitate inflow of gas toward the
center (\cite{hunt99} and references therein). Strong bars are though to be
only a  transient phase in the life of  galaxy, and form relatively quickly, 
in a time that can be comparable to the dynamical timescale  at the end of the
bar of UGC 3995A (\cite{combes93}).  These results are consistent with our
suggestion that the encounter  is recent.

In addition, recent work has shown that bars are more frequent in Seyfert 2 
than in Seyfert 1 galaxies (\cite{maialino97}). \cite{hunt99} found clear
evidence of a sequence of bar incidence that decreases from \hii/Starburst
galaxies, through LINERs, Seyfert 2 and Seyfert 1.  An important conclusion of
these authors  is that this sequence can be interpreted as an evolutionary
sequence in time, dependent also on the interaction strength (or on the
strength of any other non-axisymmetric perturbation). UGC 3995 could be seen
as an ``early'' Seyfert 2 according to this scheme.

\section{Conclusion \label{basta}} 
 
It is most likely that the  gaseous component in  the eastern part of the
disk of UGC 3995B (enhanced \ha\ emission on the side closest to UGC 3995A),
and  also in the western part of UGC 3995A (NW end of bar, region 13) has
been affected by the encounter (see \S \ref{results} and \S \ref{disc}). 
The  relative orientation, motion and morphologies of UGC~3995A and B are
most consistent if the two galaxies are not widely separated but are in the
early stages of their encounter. The presence of an active nucleus in an
early stage encounter apparently does not pose a problem, since gas in the
circumnuclear regions may reach the nucleus  on time scales comparable to the
dynamical timescale (\cite{whittle94}, and references therein). If the
encounter is prograde, then the two galaxies should be before their first
closest approach. Otherwise,  the  galaxy orientation and motions, and
especially the unperturbed \hi\ profile are consistent with a retrograde
encounter. 


\acknowledgements 
We thank W. C. Keel and D. L. Domingue for permission to measure their UGC
3995 spectrum. 
 
\newpage 
 
\begin{deluxetable}{lcccc} 
\tablewidth{0pt} 
\tablecaption{UGC~3995: Journal of Observations \label{tab:obslog}} 
\tablehead{\colhead{Date}  & \colhead{U. T.}  & 
\colhead{Exp. Time} & \colhead{Filter/Sp. Range (\AA)} 
& \colhead{P. A.} \\ 
} 
\startdata 
\\ 
\multicolumn{5}{c}{(a) Imaging} \\ 
\\ 
30-Jan-95 &      06:38 & 600 & R & -- \\ 
30-Jan-95 &      06:51 & 600 & R & --\\ 
30-Jan-95 &      07:05 & 600 & R & --\\ 
30-Jan-95 &      07:26 & 600 & H$\alpha$ & --\\ 
30-Jan-95 &      07:50 & 600 & H$\alpha$ & --\\ 
30-Jan-95 &      08:12 & 600 & H$\alpha$ & --\\ 
30-Jan-95 &      08:42 & 600 & B & --\\ 
30-Jan-95 &      08:59 & 600 & B & --\\ 
30-Jan-95 &      09:16 & 600 & B &  --\\ 
\\ 
\multicolumn{5}{c}{(b) Spectroscopy} \\ 
\\ 
9-dec-1994  &   9:22   &        1800 & 3640--6720  &    105$^\circ$ \\ 
9-dec-1994  &   9:54   &        1800 & 3640--6720  &    105$^\circ$ \\ 
9-dec-1994  &   10:36  &        1800 & 3640--6720  &     45$^\circ$    \\ 
9-dec-1994  &   11:11  &        1200 & 3640--6720  &     45$^\circ$    \\ 
2-feb-1995  &   07:49  &        1800 & 6170--7240 &     106$^\circ$ \\ 
2-feb-1995  &   08:21  &        1800 & 6170--7240 &     106$^\circ$ \\ 
2-feb-1995  &   08:53  &        1800 & 6170--7240 &     106$^\circ$ \\ 
2-feb-1995  &   09:24  &        1800 & 6170--7240 &     106$^\circ$ \\ 
2-feb-1995  &   10:07  &        1800 & 6170--7240 &     125$^\circ$ \\ 
2-feb-1995  &   10:38  &        1800 & 6170--7240 &     125$^\circ$ \\ 
3-feb-1995  &   07:45  &        1800 & 6170--7240 &     110$^\circ$ \\ 
3-feb-1995  &   08:16  &        1800 & 6170--7240 &     110$^\circ$ \\ 
3-feb-1995  &   08:53  &        1800 & 6170--7240 &     45$^\circ$  \\ 
3-feb-1995  &   09:25  &        1800 & 6170--7240 &     45$^\circ$  \\ 
4-feb-1995  &   07:22  &        1800 & 6170--7240 &     106$^\circ$ \\ 
4-feb-1995  &   07:54  &        1800 & 6170--7240 &     106$^\circ$ \\ 
4-feb-1995  &   08:38  &        1800 & 6170--7240 &     45$^\circ$  \\ 
\tablecomments{--: not available or not applicable.}
\enddata 
\end{deluxetable}

\begin{deluxetable}{rrr} 
\tablecaption{Photometric Measurements and Models \label{tab:phot}} 
\tablewidth{22pc} 
\tablehead{ 
\colhead{Parameter}& UGC~3995B & UGC~3995A} 
\startdata 
& \multicolumn{2}{c}{Measurements} \\ 
\nl 
$\rm B_T$&   \multicolumn{2}{c}{12.5}        \\ 
$\rm R_T$&   \multicolumn{2}{c}{11.4}  \\ 
$\rm \langle B-R \rangle$   &   \multicolumn{2}{c}{1.2}  \\ 
\nl 
& \multicolumn{2}{c}{Model I} \\ 
\nl 
$\rm \mu_0$         &   $21.00\pm0.05$ &  $19.80\pm0.05$ \\ 
$\rm r_h$           &   $12.79\pm0.64$ &  $19.30\pm0.96$ \\ 
$\rm \mu_e$         &   $22.50\pm0.05$ &  $19.20\pm0.05$ \\ 
$\rm r_e$           &   $1.88\pm0.09$  &  $2.66\pm0.13$ \\ 
$\rm (b/a)_{bulge}$ &   $1.00\pm0.05$  &  $0.89\pm0.05$ \\ 
$\rm (b/a)_{disk}$  &   $1.00\pm0.05$  &  $0.51\pm0.05$ \\ 
P.A.                &   $0.0\pm0.10$   &  $102.7\pm0.10$ \\ 
n                   &   $1.0\pm0.05$   &  $0.80\pm0.05$ \\ 
B$_{\rm disk} $     &   $13.47\pm0.13$ &  $12.10\pm0.16$ \\ 
B$_{\rm bulge}$     &   $18.45\pm0.13$ &  $14.60\pm0.14$ \\ 
B/D                 &   $0.01\pm0.18$  &  $0.10\pm0.21$ \\ 
B$_{\rm _tot} $     &   $13.46\pm0.18$ &  $12.00\pm0.21$ \\ 
B$_{\rm A+B}$           & \multicolumn{2}{c}{$11.75\pm0.21$} \\ 
\nl 
& \multicolumn{2}{c}{Model II} \\ 
\nl 
$\rm \mu_0$         &   $21.62\pm0.05$ &  $20.82\pm0.05$ \\ 
$\rm r_h$           &   $12.78\pm0.64$ &  $18.98\pm0.95$ \\ 
$\rm \mu_e$         &   $23.10\pm0.05$ &  $19.86\pm0.05$ \\ 
$\rm r_e$           &   $1.95\pm0.09$  &  $2.99\pm0.15$ \\ 
$\rm (b/a)_{bulge}$ &   $1.00\pm0.05$  &  $0.89\pm0.05$ \\ 
$\rm (b/a)_{disk}$  &   $1.00\pm0.05$  &  $0.51\pm0.05$ \\ 
P. A.               &   $0.0\pm0.10$   &  $102.7\pm0.10$ \\ 
n                   &   $1.0\pm0.05$   &  $0.80\pm0.05$ \\ 
B$_{\rm disk} $     &   $14.09\pm0.13$ &  $13.16\pm0.13$ \\ 
B$_{\rm bulge}$     &   $18.95\pm0.13$ &  $15.01\pm0.14$ \\ 
B/D                 &   $0.01\pm0.18$  &  $0.18\pm0.21$ \\ 
B$_{\rm tot}$       &   $14.08\pm0.18$ &  $12.98\pm0.21$ \\ 
B$_{\rm A+B}$           & \multicolumn{2}{c}{$12.64\pm0.21$} \\ 
\enddata 
\end{deluxetable}

\begin{deluxetable}{ccccl} 
\tablecaption{Narrow Band Photometry: \ha +\nii\ Fluxes\tablenotemark{a} \label{tab:ha}} 
\tablehead{\colhead{Id.} & 
\multicolumn{2}{c}{Angular Distance\tablenotemark{b}} 
& \colhead{Flux} & \colhead{Description}\\ 
 & \colhead{$\Delta\alpha$} & \colhead{$\Delta\delta$} & & } 
\startdata 
1       &  35   &  22   &        5.6  & Bright \hii\ on tip of NE spiral arm \\ 
2       &  34   &  -4   &           4.9  & Two faint \hii\ regions \\ 
3       &  \nodata &\nodata&            14.3 & Bright double \hii\ region \\ 
3a       &  33  & -12   &       \nodata &  \\ 
3b      &  30   & -15   &       \nodata  &                              \\ 
4       &  27   &   4   &           1.5  & Faint \hii\ region      \\ 
5       & \nodata & \nodata &      14.3 & Inner Southern spiral arm \\ 
5a      &  20   &   0   &       \nodata &                       \\ 
5b      &  19   &  -6   &       \nodata &                       \\ 
5c      &  10   &  -11  &       \nodata &                       \\ 
5d      &   3   &  -13  &       \nodata &                       \\ 
6       & \nodata & \nodata & 14.9 & Diffuse emission from SE spiral arm \\ 
6a      &  17   &  -21  &         \nodata &  \\ 
6b      &  12   &  -22  &         \nodata       &                       \\ 
7       &   0   &   0   &           112   & Nucleus UGC~3995A          \\ 
8       &   -5  &   22  &           3.4 & \hii\ region on northern arm \\ 
9       & \nodata&\nodata&      12.8    & Two \hii\ region, one faint \\ 
9a      &   -9  &   14  &       \nodata        &  \\ 
9b      &  -10  &  16   &       \nodata        &  \\ 
10      & -13   &   4   &           2.9 & \hii\ region nearly aligned with the nuclei of UGC~3995A and B\\ 
11      & -16   &  0    &           2.7 & Bright \hii\ region on the pseudo-ring\\ 
12      & -20   &  18   &           7.9 & Bright \hii\ region on northern arm \\ 
13      & -20   &  9    &           10.4 & Bright knot at center of pseudo-ring \\ 
14      &  -23  &  3    &           3.2   & \hii\ region on south arc of the pseudo-ring\\ 
15      &  -28  &  6    &           7.1 & Nucleus of UGC~3995B \\ 
16      &  -28  &  -3   &           6.9   & Bright \hii\ region on south arc of the pseudo-ring\\ 
17      & -31   &  12   &           3.7 & \hii\ region \\ 
18      & -35   &   5   &          4.5 & Faint \hii\ regions close to UGC~3995B\\ 
19      & -44   &   8   &           2.1 & Faint \hii\ region on western arm of UGC~3995B\\ 
--      &  --   &  --   &           572   & Total \ha\ + \nii\ flux \\ 
\enddata 
\tablenotetext{a}{In units of 10$^{-15}$ ergs s$^{-1}$. Fluxes are uncorrected for internal 
absorption, and for galactic extinction: for $\rm A_B \approx 0.09$, the 
resulting \ha\ + \nii\ fluxes should be higher by 5 \%, an increment that is 
smaller than the uncertainties associated to the calibration.} 
\tablenotetext{b}{In seconds of arcs, measured from the \ha\ centroid of the 
Seyfert nucleus (region 7); $\Delta\alpha$ \ and $\Delta\delta$ \ are 
positive to the east and to the north respectively.} 
\tablecomments{\nodata: not measured; not applicable} 
\end{deluxetable}

\begin{deluxetable}{llllll} 
\tablecaption{Emission Lines Fluxes and Widths \label{tab:eml}} 
\tablehead{ \colhead{Line Id.} & \multicolumn{2}{c}{UGC~3995A} && 
\multicolumn{2}{c}{UGC~3995B}\\ \cline{2-3} \cline{5-6} 
& \colhead{Flux} & \colhead{FWHM\tablenotemark{a}} && \colhead{Flux} & 
\colhead{FWHM\tablenotemark{a}}} 
\startdata 
\heii                   &  2.7:         &  \nodata      &&  \nodata & \nodata \\ \hb\                    &  6.3          &  350          &&  \nodata & \nodata\tablenotemark{b} \\
\o4959                  &  19.1         &  380          &&  \nodata & \nodata \\ \oiiis                   &  55.8         &  340          &&  \nodata & \nodata
\tablenotemark{b} \\ 
{\sc [Nii]}$\lambda$6548 &  \nodata      &  \nodata      &&  \nodata & \nodata   \\ 
\ha\                    &  23.22        &  370          &&  1.4        & $\simlt$100    \\ 
{\sc [Nii]}$\lambda$6583 &  22.93        &  370          &&  0.6     & $\simlt$100       \\ 
{\sc [Sii]}$\lambda$6716 &  6.8          &  320          &&  0.25:   & $\simlt$100 
\\ 
{\sc [Sii]}$\lambda$6730 &  6.4          &  330          &&  0.22:   &  \nodata \\ 
\enddata 
\tablenotetext{a}{FWHM values have been corrected for instrumental 
width, assuming FWHM$\rm _{instr}$ $\approx$ 410 \kms, for the \hb\ and \oiii\ lines, and 
$\approx$ 100 \kms\ for the \ha\ and \nii\ lines from the measurement of faint 
comparison and sky lines. 
\tablenotetext{b}{\oiiis/\hb $\approx$ 0.21}
} 
\tablecomments{\nodata: not revealed or not measured due to poor quality. ``:'': highly uncertain.} 
\end{deluxetable}

\newpage 
 
\begin{thebibliography}{} 
\bibitem[Begelman 1994]{beg94} Begelman, M. C., 1994 in Mass-Transfer Induced 
Activity in Galaxies, (Cambridge: CUP), I. Shosman Ed., 23 
\bibitem[Binney \&\ Tremaine 1987]{binney} Binney, J. \&\ Tremaine, S., 1987, 
Galactic Dynamics (Princeton: Princeton University Press) 
\bibitem[Blandford 1990]{bland90} Blandford, R. D. 1990, in Active Galactic 
Nuclei, (Berlin: Springer) Saas-Fee Advanced Course 20, 161 
\bibitem[Bressan et al. 1996]{bressan96} Bressan, A., Chiosi, C., \&\ Tantalo 
R. 1996, A\&A, 311, 425 
\bibitem[Byun 1995]{Byun} Byun, Y. I. \&\ Freeman, K. C.  1995, ApJ 448, 563 
\bibitem[Caon, Capaccioli \&\ D'Onofrio 1993]{CCD93}Caon N., Capaccioli M., \&\ 
D'Onofrio M. 1993, MNRAS, 265, 1013 
\bibitem[Chatzichristou, Vanderriest \&\ Lehnert 1998]{CVL98} 
Chatzichristou, E. T., Vanderriest C. \&\ Lehnert, M. 1998. A\&A 330, 841 
\bibitem[Combes \& Elmegreen 1993]{combes93} Combes, F., \& Elmegreen, D. M.
1983 A\&A, 271, 391
\bibitem[Combes 1997]{combes97} Combes, F. 1997, IAU Symposia, 186, in press 
\bibitem[De Robertis, Yee \& Hayhoe 1998]{der98} De Robertis, 
M. M., Yee, H. K. C. \& Hayhoe, K. 1998, \apj, 496, 93 
\bibitem[de Vaucouleurs 1959]{deV59} de Vaucouleurs, G. 1959, AL 64, 397 
\bibitem[de Vaucouleurs et al.\ 1991]{RC3} de Vaucouleurs, G., de Vaucouleurs,A., Corwin, H. G., Jr., Buta, R. J., Paturel, 
G., Fouqu\'e, P. 1991, Third Reference Catalogue of Bright Galaxies (New York: Springer) 
\bibitem[Domingue, Keel \&\ White 1998]{domin98} Domingue, D., L., Keel, W. C., \&\ White, R. E. III 1998, ApJ, in press
\bibitem[D'Onofrio et al. 1999]{d98} D'Onofrio et al. 1999, in preparation 
\bibitem[Dultzin-Hacyan,  Marziani \&\ D'Onofrio 1998]{iau186} 
Dultzin-Hacyan, D., Marziani, P., \&\ D' Onofrio, M. 1998, IAU Symposia, 186, in press
\bibitem[Dultzin-Hacyan et al. 1999a]{iau186int} Dultzin-Hacyan, 
D., Fuentes, I., Krongold, Y., \&\ Marziani, P. 1999a,  IAU Symposia, 186, in press 
\bibitem[Dultzin-Hacyan et al. 1999b]{deb98a} Dultzin-Hacyan, 
D., Fuentes, I., Krongold, Y., \&\ Marziani, P. 1999b, in Structure and 
Kinematics of Quasar Broad Line Regions, Proceeding of a meeting held in 
Lincoln on March 23-26, 1998, in press. 
\bibitem[Dultzin-Hacyan et al. 1999]{deb98b} Dultzin-Hacyan et al., 1999, in preparation.
\bibitem[Hodge 1975a]{hodge75a} Hodge, P. W. 1975a, ApJ 201, 556 
\bibitem[Hodge 1975b]{hodge75b} Hodge, P. W. 1975b, ApJ 202, 619 
\bibitem[Hodge \&\ Kennicutt 1983]{hodge83} Hodge, P. W. \&\ 
Kennicutt, R. C., J. 1983,  \aj, 88, 296 
\bibitem[Holmberg 1958]{holm58} Holmberg, E. 1958, Medd. Lund Ser. II, No. 136 
\bibitem[Hunt \& Malkan 1999]{hunt99} Hunt, L. K., \&\ Malkan, M. A. M. 1999,
ApJ, in press (SISSA preprint number astro-ph/9901410) 
\bibitem[James \&\ Puxley 1993]{puzz93} James, P. A., \&\ Puxley, P. J. 1993, 
Nature 363, 240 
\bibitem[Karachentseva 1973]{kara72} Karachentseva, V. E. 
1973, The catalogue of isolated galaxies., Soobshcheniya Spetsial'noj 
Astrofizicheskoj Observatorii, 8, 3 
\bibitem[Karachentseva et al. 1988]{kara76} 
Karachentseva V.E., Karachentsev I.D., Lebedev V.S. 1988, Astrof. Issledovanija 
Byu. Spec. Ast. Obs., 26, 42 
\bibitem[Keel 1985]{keel85} Keel, W. C. 1985, AJ 90, 2207 
\bibitem[Keel 1996]{keel96} Keel, W. C. 1996, AJ 111, 696 
\bibitem[Kennicutt, Edgar, \&\ Hodge 1989]{kenn89} Kennicutt, R. 
C., J., Edgar, B. K., \&\ Hodge, P. W. 1989,  \apj, 
337, 761 
\bibitem[Kennicutt 1992]{ken92} Kennicutt, R. C. 1992, ApJS 79, 255 
\bibitem[Laurikainen \&\ Salo 1995]{fins95} Laurikainen, E., \&\ Salo, H. 1995, 
A\&A, 293, 683 
\bibitem[Laurikainen, Salo, \&\ Aparicio 1998]{fins98} Laurikainen, E.,  Salo,  
H. \&\ Aparicio, A. 1998, A\&AS 129, 517 
\bibitem[Maiolino et al. 1997]{maialino97}
Maiolino, R., Ruiz, M., Rieke, G. H., \& Papadopoulos, P. 1997, ApJ, 485,
552
\bibitem[Malkan, Gorjian, \&\ Tam 1998]{malkan98} Malkan, M. A., 
Gorjian, V., \&\ Tam, R. 1998, \apjs, 117, 25 
\bibitem[Marquez \&\ Moles 1994]{marquez94} Marquez, I., \&\ Moles, M. 1994, 
AJ 108, 90 
\bibitem[Marziani 1995]{marz95} Marziani, P. 1995, IA-UNAM Internal Report 
CI-95-04 
\bibitem[Odewahn et al. 1992]{ode92} Odewahn, S. C., 
 Bryja, C., Humphreys, R. M. 1992, PASP 104, 553 
 \bibitem[Osterbrock 1993]{ost93} Osterbrock, D. E. 1993, ApJ, 404, 551 
\bibitem[Pronik \&\ Metik 1990]{pron90} Pronik, I. I. \&\ Metik, 
L. 1990, , NASA, Marshall Space Flight Center, Paired and Interacting Galaxies: 
IAU Colloquium No. 124,  331 
\bibitem[Rafanelli \&\ Marziani 1992]{pirlaf1}  
Rafanelli, P., \&\ Marziani, P.,  1992, AJ 103, 743 
\bibitem[Rubin et al. 1980]{rubin80} Rubin, V. C., Ford, W. K., \&\ Thonnard, 
N. 1980, ApJ 238, 471 
\bibitem[Ryder \&\ Dopita 1993]{ryder93} Ryder, S. D. \&\ Dopita, M. A. 1993, ApJS, 88, 415 
\bibitem[Schweizer \&\ Thonnard 1985]{schw85} Schweizer, F., \&\ Thonnard, N. 1985, PASP, 97, 104 
\bibitem[Stockton 1999]{stock99} Stockton, A.  1999, IAU 
Symposia, 186, in press.
\bibitem[Sulentic \&\ Arp 1983]{sul83} Sulentic, J. W. \&\ Arp, H. 1983, AJ, 89, 489 
\bibitem[Toomre \&\ Toomre 1972]{toom72} Toomre, A., \&\ Toomre, J. 1972, ApJ 
178, 623 
\bibitem[Valentijn 1990]{vale90} Valentijn, E. A. 1990, Nature 346, 153 
\bibitem[White \&\ Keel 1992]{wk92}White, R. E., \&\ Keel, W. C. 1992, Nature, 
359, 129 
\bibitem[Whittle 1994]{whittle94} Whittle, M, 1994 in Mass-Transfer Induced 
Activity in Galaxies, (Cambridge: CUP), I. Shosman Ed., 63 
\end{thebibliography}
\end{document}